\documentstyle[12pt]{article}
\newcommand{\be}{\begin{equation}}
\newcommand{\ee}{\end{equation}}
\newcommand{\ben}{\begin{eqnarray}}
\newcommand{\een}{\end{eqnarray}}
\begin{document}

\begin{center}
{\bf Topological Twistons in Crystalline Polyethylene}\footnote{This
work is partially supported by the U. S. Department of Energy (D.O.E.)
under cooperative research agreement DE-FC02-94ER40818, and by the
brazilian agencies CAPES and CNPq.}
\end{center}

\begin{center}
D. Bazeia\footnote{On leave from Departamento de F\'\i sica, Universidade
Federal da Para\'\i ba, Caixa Postal 5008, 58051-970, Jo\~ao Pessoa,
Para\'\i ba, Brazil} and E. Ventura\footnote{On leave from Departamento de
Qu\'\i mica Fundamental, Universidade Federal de Pernambuco, 50670-901,
Recife, Pernambuco, Brazil}
\end{center}

\begin{center}
Center for Theoretical Physics\\
Laboratory for Nuclear Science and Department of Physics\\
Massachusetts Institute of Technology, Cambridge, Massachusetts 02139-4307
\end{center}

\vskip 1.0cm

\begin{center}
(MIT-CTP-2790)
\end{center}

\vskip 1.5cm

\begin{center}
Abstract
\end{center}

We introduce an alternate model to describe twistons in crystalline
polyethylene. The model couples torsional and longitudinal degrees of
freedom and appears as an extension of a model that describes only the
torsional motion. We find exact solutions that describe stable topological
twistons, in good agreement with the torsional and longitudinal
interactions in polyethylene.

\newpage

The existence of twistons in polyethylene (PE) crystals was postulated
\cite{mbo78} two decades ago, and refers to a twist of 180$^0$
that extends smoothly over several $CH_2$ groups in crystalline PE, in the
plane orthogonal to the chain direction, with the corresponding
$CH_2$ unit length contraction along the polymeric chain. These twiston
configurations appear in cristalline PE as a result of its
large torsional flexibility, a general feature of unsatured polymers
which has been shown to be of widespread interest \cite{boy85,zzo92,zze92}.
In crystalline PE the presence of topological twistons may contribute
to elucidate some of its properties, in particular the dieletric
$\alpha$ relaxation \cite{wil72}.

There are some interesting models of twistons in crystalline PE
\cite{man80,swo80,spa84,zco93,zco94}. The works \cite{man80,swo80}
are almost simultaneous to the work \cite{ssh79}, which introduces
solitons to describe conductivity in polyacetylene (PA) via distortions of the
single-double bond alternations. We recall that PA engenders 
a theoretical formulation, which can follow both the
condensed matter and the field theoretic formulations \cite{jsc81}. Within
the field theoretic formulation PA seems to engenders
the most interesting realization of the phenomenon of fermion number
fractionalization -- see for instance Ref.~{\cite{jac95}}. The PA
story shows that the relevant degrees of freedom can be described
by a single real scalar field, coupled to fermionic field. The scalar
field describes longitudinal motion in the PA chain and contains
a $\phi^4$ potential to describe the double degeneracy due to the
single-douple bonds alternations in the polymeric chain. In this case
the fermionic field in introduced to describe electronic delocalization
in the polymeric chain. In the PE chain the situation is different since
there is only single bounds in the chain and so no electronic delocalization
is present anymore. This leads to descriptions envolving no fermionic degrees
of freedom, but now coupling bosonic degrees of freedom to describe the
torsional flexibility of this unsatured polymer according to the
model under consideration \cite{man80,swo80,spa84,zco93,zco94}.

In the present work we consider the crystalline PE chain as a planar
zig-zag chain with the molecular group $CH_2$ in such a way that the distance
between two consecutive groups of $CH_2$ is $2r_0$ ($r_0=0.42\,{\rm\AA}$) in
the direction orthogonal to the chain and $c$ ($c=1.27\,{\rm\AA}$) in the
direction along the chain. We obtain degenerate ground states by performing a
$180^o$ rotation in the left or right sense of the (torsional) direction
orthogonal to the chain length, or by performing a tranlation by $c$ in the
positive or negative sense of the (longitudinal) direction along the chain.
We shall offer an alternate model that couples torsional and longitudinal
degrees of freedom in a way such that the model now supports exact
twiston-like solutions, stable and in good agreement with
properties of crystalline PE.

The pioneer works \cite{man80,swo80} have considered
approximations for the interactions that
ultimately result in models describing the twiston degree of freedom
uncoupled from the longitudinal motion of the $CH_2$ molecular group --
see also \cite{spa84}. In the more recent models \cite{zco93,zco94}
one includes interactions between radial,
torsional and longitudinal degrees of freedom. In this case one uses
cilindrical coordinates to describe a generic $CH_2$ unit via
$(r_n,\theta_n,z_n)$, which corresponds to the three degrees of
freedom of the rigid molecular group. A simplification can be introduced,
and concerns freezing the $r_n$'s, so that the radial motion is neglected.
In \cite{zco93} one further ignores the translational
degree of freedom, the $z_n$ coordinates, to get to a simple model described
via the torsional variable that in the continuum limit can be taken as
$\theta_n(t)\to\theta(z,t)$. The model reproduces the double sine-Gordon model,
according to the assuptions there considered to describe the intermolecular
interaction.

The other more recent work \cite{zco94} on twiston in crystalline PE gives
another step toward a more realistic model. In this case the three
coordinates $(r_n,\theta_n,z_n)$ are taken into account to describe twistons
in PE crystals. This is the first time the radial,
torsional and longitudinal degrees of freedom are simultaneously
considered to model twiston in crystalline PE. The model is
interesting, although it is hard to find exact solutions and investigate
the corresponding issues of stability. The problem engenders several
intrinsec difficulties, which have inspired us to search for an alternate
model, in the form of two coupled fields belonging to the class of systems
investigated in the recent works \cite{bsr95,bsa96,brs96,bnt97}.

The basic assumptions that appear in former
models for twistons in crystalline PE may be introduced by considering
$(r_n,\theta_n,z_n)$ to describe degrees of freedom related
to a generic $CH_2$ molecular group in the polymeric chain. The Lagrangian
presents the usual form
\ben
L&=&\,T\,-\,U
\\
\label{ke}
T&=&\frac{1}{2}\,m\,\sum_{n}\,({\dot{r}}_n^2+r_n^2\,{\dot{\theta}}_n^2+
{\dot{z}}_n^2)
\\
U&=&U_{\rm intra}+U_{\rm inter}
\een
where $m$ is the mass associated to the molecular group $CH_2$, and
$U_{\rm intra}$ and $U_{\rm inter}$ are potentials, the first related
to the intramolecular interactions, and the second used to model the
intermolecular forces in the crystalline environment. The intramolecular
potential can be considered as
\be
\label{intra}
U_{\rm intra}=\frac{1}{2}\,\sum_n\,K_1\,(\theta_{n+1}-\theta_n)^2+
\frac{1}{2}\,\sum_n\,K_2\,(z_{n+1}-z_n)^2+\cdots
\ee
where $K_1$ and $K_2$ are coeficients related to the harmonic
approximation for torsional and longitudinal motions, respectively. The
intramolecular potential may contain derivative coupling between the
torsional and longitudinal motions \cite{zco93}, or between the radial
and longitudinal motions \cite{zco94}. We shall not consider such
possibilities in the present work, although in \cite{bnt97} one shows a
route for taking derivative coupling into account. The second potential
responds for the intermolecular interactions and is usually given in the
form
\be
\label{inter}
U_{\rm inter}=\sum_n\,[U_0(\theta_n)+U_1(\theta_n)\,U_l(z_n)]
\ee
where $U_0(\theta_n)$ and $U_1(\theta_n)$ are used to model torsion and
$U_l(z_n)$ to describe the longitudinal motion along the chain. In the
works \cite{man80,swo80,spa84,zco93}, after freezing radial and translational
motion, the above intermolecular potential is described by the
$U_0(\theta_n)$ contributions. Here one gets to a system described by the 
torsional motion alone, and in the continuum limit there are some models,
described via the sine-Gordon potential \cite{man80,swo80,spa84}
\be
\label{sg}
U_0(\theta)=A_1\,[1-\cos(2\,\theta)]
\ee
or the polynomial potential \cite{swo80}
\be
U_0(\theta)=-\frac{1}{2}\,A_2\,\theta^2+\frac{1}{4}\,B_1\,\theta^4
\ee
or yet the double sine-Gordon potential \cite{zco93}
\be
\label{v0}
U_0(\theta)=A_3\,[1-\cos(2\,\theta)]+B_2\,[1-\cos(4\,\theta)]
\ee
The $A_i$ and $B_i$ are real constants, used to parametrize the
corresponding interactions.

The more recent work \cite{zco94} considers coupling between the radial,
torsional and longitudinal degrees of freedom. There the intermolecular
potential was written as
\be
U_{inter}=U_0(\theta)+B_1\sin(\theta-51^o)\cos\left(\pi\frac{u}{c}\right)
\ee
where $u(z,t)$ is the continuum version of $u_n(t)=z_n(t)-nc$. Here
$U_0(\theta)$ is given in $(\ref{v0})$ and we have neglected a constant,
very small phase. This model is interesting, although the analytical
solutions obtained in \cite{zco94} are found under assumptions that
ultimately decouple the system.

The difficulties inherent to this problem bring motivations for
introducing other assumptions, with the aim of offering an alternate model
that presents exact solutions for twistons in crystalline PE. Toward this
goal, let us use cilindrical coordinates to describe the molecular groups
under the assumption of rigidity. We start with the kinetic energy
$(\ref{ke})$, rewriting it in the form
\be
T=\frac{1}{2}\,m\,r_0^2\sum_{n}\left({\dot{\phi}}_n^2+
\left(\frac{c}{r_0}\right)^2\,{\dot{\chi}}_n^2+\dot{\rho}_n^2\right)
\ee
Here we have set $\phi_n=\theta_n-[1-(-1)^n](\pi/2)$, $\chi_n=(z_n-nc)/c$
and $\rho_n=(r_n-r_0)/r_0$, where $r_0$ is the equilibrium position
of the radial coordinate and $c$ in the longitudinal
distance between consecutive molecular groups. Now $\phi_n$, $\chi_n$ and
$\rho_n$ are all dimensionless variables, and in the continuum
limit they can be seen as real fields $\phi(z,t)$, $\chi(z,t)$ and
$\rho(z,t)$. Before going to the continuum version of the PE chain,
however, let us reconsider the intramolecular potential given by
Eq.~(\ref{intra}). We use the harmonic approximation to write
\be
\label{intrag}
U_{\rm intra}=\frac{1}{2}\,\sum_n\,k_t\,(\phi_{n+1}-\phi_n)^2+
\frac{1}{2}\,\sum_n\,k_l\,(\chi_{n+1}-\chi_n)^2+
\frac{1}{2}\,\sum_n\,k_r\,(\rho_{n+1}-\rho_n)^2
\ee
where $k_t$, $k_l$ and $k_r$ are spring-like constants, related to the
torsional, longitudinal and radial degrees of freedom, respectively. 

The harmonic interactions present in the
intramolecular term $(\ref{intrag})$ makes the dynamics to appear as the
dynamics of relativistic fields, in the same way it happens with the standard
harmonic chain. For this reason the continuum version
of our general model can be written in the form
\ben
{\cal L}_3&=&\frac{1}{2}\left(\frac{\partial\phi}{\partial t}\right)^2-
\frac{1}{2}\left(\frac{\partial\phi}{\partial z}\right)^2+
\frac{1}{2}\left(\frac{\partial\chi}{\partial t}\right)^2-
\frac{1}{2}\left(\frac{\partial\chi}{\partial z}\right)^2+\nonumber\\
& &\frac{1}{2}\left(\frac{\partial\rho}{\partial t}\right)^2-
\frac{1}{2}\left(\frac{\partial\rho}{\partial z}\right)^2
-V(\phi,\chi,\rho)
\een
This is a model describing three real scalar fields and we are working
in the bidimensional spacetime, using natural units $(\hbar=c=1)$ and standard
relativistic notation.

The general model presents the same coeficient in all the derivative terms
and the potential is used to specify the particular model one is interested
in. This is a general feature of field theoretic systems and specific models
may support topological solitons when the potential presents discrete symmetry
and at least two degenerate minima \cite{jac77,raj82}. We remark that system
like the above one has been recently considered  \cite{chw98} as an effort
to understand issues of strings ending on walls, in this case bringing the
concept of complete wetting from condensed matter to field theory. In the
present work, however, the potential is supposed to have the form
\be
V(\phi,\chi,\rho)=\frac{1}{2}\left(\frac{\partial H}{\partial\phi}\right)^2+
\frac{1}{2}\left(\frac{\partial H}{\partial\chi}\right)^2+
\frac{1}{2}\left(\frac{\partial H}{\partial\rho}\right)^2
\ee
where $H=H(\phi,\chi,\rho)$ is a smooth but otherwise arbitrary
function of the fields. This restriction is introduced along the lines of
former investigations \cite{bsr95,bsa96,brs96}, and leads to interesting
properties. For instance, the equations of motion for static configurations
$\phi=\phi(z)$, $\chi=\chi(z)$ and $\rho=\rho(z)$ present the form
\ben
\frac{d^2\phi}{dz^2}&=&H_{\phi}H_{\phi\phi}+H_{\chi}H_{\chi\phi}+
H_{\rho}H_{\rho\phi}
\\
\frac{d^2\chi}{dz^2}&=&H_{\phi}H_{\phi\chi}+H_{\chi}H_{\chi\chi}+
H_{\rho}H_{\rho\chi}
\\
\frac{d^2\rho}{dz^2}&=&H_{\phi}H_{\phi\rho}+H_{\chi}H_{\chi\rho}+
H_{\rho}H_{\rho\rho}
\een
where $H_{\phi}$ stands for $\partial H/\partial\phi$, and so on. These
equations are solved by the following first-order
differential equations \cite{bsr95,brs96}
\ben
\frac{d\phi}{dz}&=&H_{\phi}
\\
\frac{d\chi}{dz}&=&H_{\chi}
\\
\frac{d\rho}{dz}&=&H_{\rho}
\een
Solutions that solve these first-order equations are classically or linearly
stable \cite{bsa96,brs96}.

In the present work, however, our main interest is to describe twistons in
crystalline PE. For this reason we now focus attention on this specific
polymeric chain. As one knows, in this case it is a very good approximation
\cite{man80,swo80,spa84,zco93,zco94} to descard radial motion in the PE chain.
This simplification leads to a system of two fields, describing torsional and
longitudinal motions simultaneously. However, we give up the route introduced
in former models and introduce a new way, obtained after getting
inspirations from field theoretic analysis \cite{jsc81,jac77,raj82} and
from a former model of two coupled fields, used to model topological solitons
in ferroelectric crystals \cite{brs96}. For simplicity, however, we first
think of a system describing the torsional motion {\it alone}. In this case
we get the Lagrangian density
\be
{\cal L}_1=\frac{1}{2}\left(\frac{\partial\phi}{\partial t}\right)^2-
\frac{1}{2}\left(\frac{\partial\phi}{\partial z}\right)^2-V_1(\phi)
\ee
We now seek for a system that is in good approximation to realistic
torsional interactions in PE. According to Refs.~\cite{zco93,zco94},
investigations on molecular simulation allows introducing the following
torsional potential
\be
\label{p1}
V_1(\phi)=\frac{1}{2}\lambda^2\phi^2(\phi^2-\pi^2)^2
\ee
which is generated via the function
\be
H_1(\phi)=\frac{1}{2}\,\lambda\,\phi^2\,(\frac{1}{2}\phi^2-\pi^2)
\ee
This potential presents three minima, one at $\phi=0$ and the other
two at $\phi^2=\pi^2$, corresponding to a $180^o$ rotation in the left or right
sense of the torsional direction, in agreement with the degenerate ground
states of crystalline PE.

In this case the equation of motion for static configuration is
\be
\frac{d^2\phi}{dz^2}=\lambda^2\phi(\phi^2-\pi^2)(3\phi^2-\pi^2)
\ee
It is solved by solutions of the first-order equation
\be
\frac{d\phi}{dz}=\lambda\phi(\phi^2-\pi^2)
\ee
There are topological twistons, given by \cite{bsr95}
\be
\label{s1}
\phi^{(\pm)}_{(t)}(z)=\pm\,\pi\,
\sqrt{\frac{1}{2}[1-\tanh(\lambda \pi^2 z)]\,}
\ee
Here we are taking $z=0$ as the center of the soliton, but this is unimportant
because the continuum, infinity chain presents translational invariance.
The sign of $\lambda$ identifies kink and antikink solutions,
connecting the minima $0$ and $\pi$ or $0$ and $-\pi$.
These solutions are stable and can be boosted to their time-dependent
form by just changing $z$ to $\xi=(z-vt)/(1-v^2)^{1/2}$. This model can be
seem as an alternate model to the ones previously introduced to describe the
torsional motion alone \cite{man80,swo80,spa84,zco93}.

To get to a more realistic model we couple the torsional field to the
longitudinal motion along the chain. We model the presence of interactions
by extending the former function $H_1(\phi)$ to $H_2(\phi,\chi)$ given by
\be
\label{h2}
H_2(\phi,\chi)=\frac{1}{2}\,\lambda\,\phi^2\,(\frac{1}{2}\phi^2-\pi^2)+
\frac{1}{2}\mu\phi^2\chi^2
\ee
In this case we get the potential
\be
\label{p2}
V_2(\phi,\chi)=\frac{1}{2}\lambda^2\phi^2(\phi^2-\pi^2)^2+
\lambda\mu\phi^2(\phi^2-\pi^2)\chi^2+\frac{1}{2}\mu^2\phi^2\chi^4+
\frac{1}{2}\mu^2\phi^4\chi^2
\ee
Among the several features of this potential we remark that
\be
V_2(\phi,0)=V_1(\phi)
\ee
which reproduces the torsional model $V_1(\phi)$ when one
freezes the longitudinal motion. Also,
\ben
V_2(0,\chi)&=&0
\\
V_2(\pm\pi,\chi)&=&\frac{1}{2}\mu^2\pi^4\chi^2+\frac{1}{2}\mu^2\pi^2\chi^4
\een
We can evaluate the quantity $\partial^2 V/\partial\phi\partial\chi$ to see
that it contributes with vanishing values at the minima $(0,0)$ and
$(\pm\pi,0)$. This shows that the spectra of excitations of the
torsional motion around the ground states are unaffected by the presence of the
longitudinal motion. Thus, we can use $V_1(\phi)$ to investigate the behavior
of the torsional motion around the equilibrium configurations. Here we get the
results that the $\phi$ field presents mass values
$m_{\phi}(0,0)=|\lambda|\pi^2$ and $m_{\phi}(\pm\pi,0)=2|\lambda|\pi^2$,
around the minima $\phi=0$ and $\phi^2=\pi^2$ respectively.
Accordingly, for the $\chi$ field we see that it is massless at $\phi=0$,
and at $\phi^2=\pi^2$ its mass becomes $m_{\chi}(\pm\pi,0)=|\mu|\pi^2$.
These results identify an asymmetry in the spectra of excitations of both
the torsional and longitudinal motion around the minima $(0,0)$ and
$(\pi^2,0)$. This asymmetry appears in consequence of the polynomial
potential $(\ref{p2})$, and is small for small
parameters $\lambda$ and $\mu$. On the other hand, these results allow
introducing the ratio $m_{\chi}/m_{\phi}$ between the masses of the $\chi$
and $\phi$ fields. At the minima $(\pm\pi,0)$ we get
$m_{\chi}/m_{\phi}=|\mu|/2|\lambda|$ and at the minimum $(0,0)$ it vanishes,
because of the flatness of the potential in the $\chi$ direction
at that point. Our model indicates that this ratio belong to the
interval $[0,|\mu|/2|\lambda|\,]$. We can very naturally identify this ratio
$m_{\chi}/m_{\phi}$ with the ratio between the longitudinal $(b_l)$ and
torsional $(b_t)$ energy barriers in the polymeric chain. Thus, if we use
the value $|\mu|/2|\lambda|$, which is the only specific value our model
provides for that quantity, we can write
\be
\label{bar}
\Big{|}\,\frac{\mu}{\lambda}\,\Big{|}=2\,\frac{b_l}{b_t}
\ee
Since the torsional and longitudinal barriers in the crystalline PE chain are
known, we can use them to evaluate the ratio between the two in principle
arbitrary parameters $\lambda$ and $\mu$ that appear in our model. We use
values presented in Fig.~[3] of Ref.~{\cite{mbo78} to see that the ratio
$b_t/b_l$ for crystalline PE is in the interval $[2.0,3.0]$. This indicates
that in our model the ratio between $\lambda$ and $\mu$ is governed by
$|\lambda|/|\mu|\in[1.0,1.5]$.

For the above model the equations of motion for static fields are given by
\ben
\frac{d^2\phi}{dz^2}&=&\lambda^2\phi(\phi^2-\pi^2)(3\phi^2-\pi^2)+
2\lambda\mu\phi(2\phi^2-\pi^2)\chi^2+\mu^2\phi(\chi^2+1)\chi^2
\\
\frac{d^2\chi}{dz^2}&=&2\lambda\mu\phi^2(\phi^2-\pi^2)\chi+2\mu^2\phi^2\chi^3+
\mu^2\phi^2\chi
\een
Although there is no general way of solving these equations, we recognize
that they follow from the potential in Eq.~{(\ref{p2})}, defined via
the function introduced in Eq.~{(\ref{h2})}, and so they are solved by
\ben
\label{foeq21}
\frac{d\phi}{dz}&=&\lambda\phi(\phi^2-\pi^2)+\mu\phi\chi^2
\\
\label{foeq22}
\frac{d\chi}{dz}&=&\mu\phi^2\chi
\een
which are first-order differential equations, easier to investigate. Here we
can find exact analytical solutions, in the form
\ben
\label{s21}
\phi_{(t,l)}^{(\pm)}(z)&=&\pm\,\pi\,\sqrt{ \frac{1}{2}[1-\tanh(\mu \pi^2z)]\, }
\\
\chi_{(t,l)}^{(\pm)}(z)&=&\pm\,\pi\,\sqrt{\frac{\lambda}{\mu}-1\,}\,
\sqrt{ \frac{1}{2}[1+\tanh(\mu\pi^2z)]\,}
\label{s22}
\een
They are valid for $\lambda/\mu>1$ and are similar to the
solutions found in Ref.~{\cite{zco94}} to describe the torsional and
longitudinal degrees of freedom that describe topological twistons in the
crystalline chain.

The solutions found in this work are stable and engender other interesting
features. For instance, for the model of a single field the width of the
solutions is inversely proportional to $|\lambda|$. For the model with two
fields the  solutions $(\ref{s21})$ and $(\ref{s22})$ present the same width,
inversely proportional to $|\mu|$. This can be understood intuitively since
we expect that when the torsional motion completes the $180^o$ rotation and
returns to crystal register, the longitudinal motion should simultaneously
return to its crystalographic position. On the other hand, this is in
agreement with the topological features of the solutions and with the
trial orbit \cite{bsr95}
\be
\phi^2+\frac{1}{\frac{\lambda}{\mu}-1}\,\chi^2=\pi^2
\ee
that was used to solve the coupled equations $(\ref{foeq21})$ and
$(\ref{foeq22})$. Furthermore, the amplitude of the torsional field is $\pi$,
as given by the solution $(\ref{s21})$, in agreement with the model
we are using for the twiston configuration. The amplitude of the longitudinal
motion can be obtained from the solution $(\ref{s22})$. It is given by
\be
\label{achi}
\pi\sqrt{\frac{\lambda}{\mu}-1\,}
\ee
In the PE chain we have to set this amplitude to unity, to make it compatible
with the fact that when the chain twists by 180$^0$ and turns to its
crystalline register the longitudinal motion shifts by $c$. This picture
follows in accordance with the fact that crystalline PE presents
degenerate ground states, obtained from each other by a rotation of 180$^0$
or by a translation of $c$ along the polymer chain. We set
the amplitude $(\ref{achi})$ to unity to get 
\be
\frac{\lambda}{\mu}=1+\frac{1}{\pi^2}
\ee
This result predicts the numerical value $\lambda/\mu=1.1$, which is in good
agreement with the reasoning that follows from the ratio between the
torsional and longitudinal barriers, just after Eq.~(\ref{bar}).

The results presented in this work add to the former investigation
\cite{brs96} to show that the approach of using systems of coupled fields
to describe topologically non-trivial excitations in continuum versions of
polymeric chains appears to work correctly. The procedure shows that some
features of the relativistic model can be directly mapped to specific
properties of the actual polymeric chain, although work on issues concerning
fine-tunning the relativistic approach is still to be done. Investigations
concerning this and other related issues are presently under consideration.

\begin{center}
Acknowledgments
\end{center}
We would like to thank Roman Jackiw for comments and for reading
the manuscript. We also thank Cl\'audio Furtado for drawing our attention to
Ref.~{\cite{zco94}} and Alfredo Simas for interest, and for discussions on
other polymeric chains. We are grateful to Robert Jaffe and the Center for
Theoretical Physics, MIT, for hospitality.


\begin{thebibliography}{20}
\bibitem{mbo78}M.L. Mansfield and R.H. Boyd, J. Polym. Sci. Phys. Ed.
{\bf 16}, 1227 (1978).

\bibitem{boy85}R.H. Boyd, Polymer {\bf 26}, 323, 1123 (1985).

\bibitem{zzo92}G. Zerbi and M.D. Zoppo, J. Chem. Soc. Faraday Trans.
{\bf 88}, 1835 (1992).

\bibitem{zze92}M.D. Zoppo and G. Zerbi, Polymer {\bf 33}, 4667 (1992).

\bibitem{wil72}G. Williams, Chem. Rev. {\bf 72}, 55 (1972).

\bibitem{man80}M.L. Mansfield, Chem. Phys. Lett. {\bf 69}, 383 (1980).

\bibitem{swo80}J.L. Skinner and P.G. Wolynes, J. Chem. Phys.
{\bf 73}, 4015, 4022 (1980).

\bibitem{spa84}J.L Skinner and Y.H. Park, Macromolecules {\bf 17}, 1735 (1984).

\bibitem{zco93}F. Zhang and M.A. Collins, Chem. Phys. Lett. {\bf 214}, 459
(1993).

\bibitem{zco94}F. Zhang and M.A. Collins, Phys. Rev. E {\bf 49}, 5804 (1994).

\bibitem{ssh79}W.P. Su, J.R. Schrieffer and A.J. Heeger, Phys. Rev.
Lett. {\bf 42}, 1698 (1979).

\bibitem{jsc81}R. Jackiw and J.R. Schrieffer, Nucl. Phys. B
{\bf 190 [SF3]}, 253 (1981).

\bibitem{jac95}R. Jackiw, {\it Diverse Topics in Theoretical and Mathematical
Physics,} (World Scientific, Singapore, 1995) pp. 79 and 449.

\bibitem{bsr95}D. Bazeia, M.J. dos Santos and R.F. Ribeiro, Phys. Lett.
A {\bf 208}, 84 (1995).

\bibitem{bsa96}D. Bazeia and M.M. Santos, Phys. Lett.
A {\bf 217}, 28 (1996).

\bibitem{brs96}D. Bazeia, R.F. Ribeiro, and M.M. Santos, Phys. Rev.
E {\bf 54}, 2943 (1996).

\bibitem{bnt97}D. Bazeia, J.R.S. Nascimento, and D. Toledo,
Phys. Lett. A {\bf 228}, 357 (1997).

\bibitem{jac77}R. Jackiw, Rev. Mod. Phys. {\bf 49}, 681 (1977).

\bibitem{raj82}R. Rajaraman, {\it Solitons and Instantons} (North-Holland,
Amsterdam, 1982).

\bibitem{chw98}A. Campos, K. Holland and U.-J. Wiese, Phys. Rev. Lett.
{\bf 81}, 2420 (1998).

\end{thebibliography}
\end{document}